\documentclass[aps,prb,reprint,groupedaddress,showpacs]{revtex4-1}
\usepackage{color}

\usepackage{graphicx}
\usepackage{mathrsfs} 
\usepackage[mathscr]{euscript}
\usepackage{amsmath,amssymb}
\usepackage{bm}

\def\Tr{{\rm Tr}}
\def\ket#1{ | #1 \rangle }
\def\bra#1{ \langle #1 | }

\def\ketbra#1#2{ \ket{#1} \bra{#2} }

\def\vec#1{ {\bm{#1}} }

\begin{document}


\title{Doubling of Entanglement Spectrum in Tensor Renormalization Group}


\author{Hiroshi \textsc{Ueda}$^1$}
\email[]{h\_ueda@riken.jp}
\author{Kouichi \textsc{Okunishi}$^2$}
\email[]{okunishi@phys.sc.niigata-u.ac.jp}
\author{Tomotoshi \textsc{Nishino}$^3$}
\email[]{nishino@kobe-u.ac.jp}
\affiliation{$^1$Condensed Matter Theory Laboratory, RIKEN, Wako, Saitama 351-0198, Japan}
\affiliation{$^2$Department of Physics, Niigata University, Niigata 950-2181, Japan}
\affiliation{$^3$Department of Physics, Graduate School of Science, Kobe University, Kobe 657-8501, Japan}

\date{\today}

\begin{abstract}
We investigate  the entanglement spectrum in HOTRG ---tensor renormalization group (RG) method combined with the higher order singular value decomposition--- for two-dimensional (2D) classical vertex models. 
In the off-critical region, it is explained that the entanglement spectrum associated with the RG transformation is described by `doubling' of the  spectrum of a corner transfer matrix. 
We then demonstrate that the doubling actually occurs for the square-lattice Ising model by HOTRG calculations up to $D = 64$, where $D$ is the cut-off dimension of tensors. 
At the critical point, we also find that a non-trivial $D$ scaling behavior appears in the entanglement entropy.
We mention about the HOTRG for the 1D quantum system as well. 
\end{abstract}


\maketitle

\section{Introduction}

The real-space renormalization group (RG) is an efficient numerical method for analysis of quantum/classical lattice systems~\cite{Kadanoff_1965_RG, Wilson_1974_RG, book:Burkhardt_1982_RSRG}. 
A main goal of the numerical RG is to extract a small number of relevant degrees of freedom, which well describe the physical properties of the ``target state'', such as a correlated groundstate or a thermal equilibrium state, in the Hilbert space of huge dimension. 
Recently a number of real-space RG formulations refer to the entanglement between the system and the environment, for the purpose of keeping relevant degrees of freedom systematically. 
A typical example is the density matrix renormalization group (DMRG), which has been a powerful computational tool for one-dimensional (1D) quantum systems and two-dimensional (2D) classical ones~\cite{White_1992_DMRG_1993_DMRG, book:Peschel_1999_DMRG, Nishino_1999_DMRG, Schollwock_2005_DMRG_Rev, Schollwock_2011_MPS_Rev}.
In DMRG, the entanglement entropy is implicitly maximized by means of the singular-value decomposition assisted by the diagonalization of the reduced density matrix~\cite{Xiang_2001_2D_DMRG,Legeza_2003_k_space_DMRG,Legeza_2004_QI_and_DMRG,Vidal_2003_TEBD}. 
It should be noted that when a gapped groundstate is targeted, the spectrum of the reduced density matrix in the bulk limit is well described by the eigenvalue
distribution of Baxter's corner transfer matrix (CTM)~\cite{book:Baxter_1982_CTM,Peschel_1999_DM_spectra, Okunishi_1999_DM_spectra}. 

As a higher dimensional extension of DMRG,  a corner transfer tensor approach  was firstly formulated for the 3D Ising model represented as a 3D vertex model~\cite{Nishino_1998_CTTRG,Orus_2012}.
This approach, however, suffers from a slow decay in eigenvalues of the reduced density matrix, where the target scheme in the RG transformation was not appropriate from the modern view point.
Complementary direct variational approaches based on the 2D tensor product state\cite{Nishino_2000_TPVA,Nishino_2001_TPVA}, or the 2D projected entangled pair state\cite{Verstraete_2006_PEPS,Jordan_2009_MPS_iPEPS,Orus_2009_CTM_iPEPS,Corboz_2010_CTM_ifPEPS,Orus_2012_review}  has been applied to the higher-dimensional problems. 
Recently, Xie {\it et al}. proposed an improved tensor RG method\cite{Levin_2007_TRG} combined with the higher order singular value  decomposition (HOSVD)~\cite{Lathauwer_2000_HOSVD}, which has been abbreviated as HOTRG~\cite{Xie_2012_HOTRG}. 
They precisely estimated the critical temperature and scaling exponents for 3D classical systems and 2D quantum systems.
Moreover, a further improvement of the tensor RG can be achieved by the second renormalization group~\cite{Xie_2009_SRG,Zhao_2010_SRG,Xie_2012_HOTRG}, but numerical cost for this improvement is relatively high.

In HOTRG, the effective tensors representing the renormalized Boltzmann weights are directly constructed by the tensor decomposition, while the second renormalization takes account of the entanglement between the system and the environment.
Thus, we may expect that these two tensor RG methods involve essentially different theoretical backgrounds in some sense.
In HOTRG, however,  the role of the entanglement has not been discussed yet, because its RG transformation does not explicitly refer to the entanglement between the system and the environment.
In this paper, we thus clarify the role of the entanglement in HOTRG for the 2D classical models, where we can refer various exact results of the integrable systems and conformal field theory(CFT).
Note that the observation of the 2D classical models is also relevant to 1D quantum systems, through the well-known quantum-classical correspondence~\cite{Suzuki_1976_trotter}. 
We also investigate the scaling of the entanglement entropy at criticality. 

This paper is organized as follows. 
In the next section, we briefly review the HOTRG for 2D vertex models on the square lattice. 
In Sec.~\ref{fp}, we discuss the structure of the renormalized vertex weight at the fixed point, on the basis of the corner double line (CDL) picture~\cite{Gu_2009_TEFR}. 
In particular, we show that the entanglement spectrum of the reduced density matrix is represented by `doubling' of the CTM spectrum.
In Sec.~\ref{des}, the numerical evidence of the CDL picture is shown in the off-critical region of the 2D classical Ising model. 
In Sec.~V, the finite size behaviors of the free energy and the entanglement entropy at the criticality are also investigated. 
In addition, we analyze the scaling of the effective correlation length with respect to the cut-off dimension $D$. 
The last section is devoted to a summary.

\section{HOTRG}
\label{gptr}

\begin{figure}[Htb]
  \centering
  \resizebox{3.4cm}{!}{\includegraphics{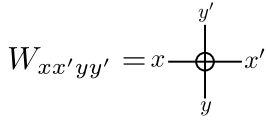}}
  \caption{Diagram for a local vertex weight, where $x,x', y, y'$ represent two states link variables.
 }
  \label{fig1}
\end{figure}

Let us consider a 2D classical vertex model on the square lattice, where a local vertex weight is represented as a 4-leg tensor
$W_{x x' y y'}^{~}$.
The indices $x$, $x'$ and $y$, $y'$ respectively correspond to link variables in the horizontal and vertical directions, as shown in Fig.~\ref{fig1}. 
Throughout this article we consider the 2-state vertex model, typically $x \in 1, 2$, etc., and assume the symmetry where $W$ is invariant under the permutation between $x$ and $x'$, and between $y$ and $y'$, for the purpose of simplifying the formulations.

\begin{figure}[Htb]
  \centering
  \resizebox{7cm}{!}{\includegraphics{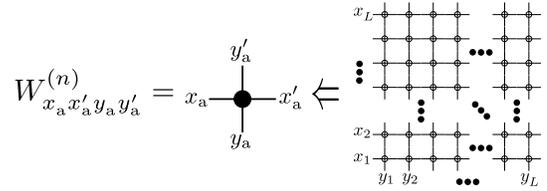}}
  \caption{ Schematic diagram of the  vertex tensor $W_{x^{~}_{\rm a} x'_{\rm a} y^{~}_{\rm a} y'_{\rm a}}^{(n)}$  on a $L\times L$ lattice with $L=2^n$. 
The indices of the tensor specify the configurations  at the edges. }
  \label{fig2}
\end{figure}

In order to capture the feature of the tensor renormalization group, we start with a system, which consists of $L \times L$ vertices connected in the geometry of a square area with a set of specified boundary configurations. 
Consider the case where the link variables at the bottom and the top boundaries of the area, respectively, are fixed as 
$y^{~}_{\rm a} \equiv \{y^{~}_1, \cdots, y^{~}_L \}$ and 
$y'_{\rm a} \equiv \{ y'_1, \cdots, y'_L \}$, 
and those at the left and the right boundaries as 
$x^{~}_{\rm a} \equiv \{ x^{~}_1, \cdots, x^{~}_L \}$ and 
$x'_{\rm a} \equiv \{ x'_1, \cdots, x'_L \}$, 
as is depicted in Fig.~\ref{fig2}. 
We put subscript ``$a$'' on the composite link variables, for the later convenience. 
Multiplying all the vertex weights in the area and taking configuration sum for those links connecting the neighboring vertices in the bulk region, we obtain the partition function 
$W_{x^{~}_{\rm a} x'_{\rm a} y^{~}_{\rm a} y'_{\rm a}}^{~}$ 
under the specified boundary configurations $x^{~}_{\rm a}$, $x'_{\rm a}$, $y^{~}_{\rm a}$, and $y'_{\rm a}$.
We have used the letter ``$W$'', rather than ``$Z$'', for the 
partition function, since it is possible to interpret 
$W_{x^{~}_{\rm a} x'_{\rm a} y^{~}_{\rm a} y'_{\rm a}}^{~}$ 
as a kind of vertex weight, where each index has $2^L_{~}$-degrees of freedom. 
Hereafter, we consider the case where the linear dimension $L$ is chosen to be $2^n_{~}$, and explicitly show the system size by putting the number on the tensor as 
$W_{x^{~}_{\rm a} x'_{\rm a} y^{~}_{\rm a} y'_{\rm a}}^{(n)}$. 

It is possible to extend the area of the lattice by joining two vertices. 
For example, let us put 
$W^{(n)}_{x^{~}_{\rm b} x'_{\rm b} y \, y'_{\rm b}}$ on top of 
$W^{(n)}_{x^{~}_{\rm a} x'_{\rm a} y^{~}_{\rm a} y}$ 
and contracting the vertical link. We then obtain a composit tensor 
\begin{equation}
M^{(n+1,n)}_{x^{~}_{\rm a} x^{~}_{\rm b} x'_{\rm a} x'_{\rm b} y^{~}_{\rm a} y'_{\rm b}} =  
\sum_y^{~} 
W^{(n)}_{x^{~}_{\rm a} x'_{\rm a} y^{~}_{\rm a} y} 
W^{(n)}_{x^{~}_{\rm b} x'_{\rm b} y \, y'_{\rm b}} \, ,
\label{joint_of_vertex}
\end{equation} 
which corresponds to the 
$2L \times L( = 2^{n+1}_{~} \times 2^{n}_{~})$ 
area on the lattice. 
Representing the index pairs 
$\{ x^{~}_{\rm a} x^{~}_{\rm b} \}$ and $\{ x'_{\rm a} x'_{\rm b} \}$, 
respectively,  as joined link variables
$x_{\rm c}^{~}$ and $x'_{\rm c}$, the tensor $M^{(n+1,n)}_{~}$ 
can also be considered as an extended vertex weight 
$M^{(n+1,n)}_{x_{\rm c}^{~} x'_{\rm c} y_{\rm a}^{~} y'_{\rm b} }$. 
In the same manner, we can align two $M^{(n+1,n)}_{~}$ horizontally and contracting the joint indices, as we have done in the above equation, we obtain a wider extended weight 
$W^{(n+1)}_{x_{\rm c}^{~} x'_{\rm c} y_{\rm d}^{~} y'_{\rm d} }$, 
which corresponds to the
$2^{n+1}_{~} \times 2^{n+1}_{~}$ 
area on the lattice. 
Thus starting from the original vertex weight $ W^{(0)}_{~}=W$, one can define $W^{(n)}_{~}$ for arbitrary $n$.

From the computational view point, there is a strong limitation on the maximal value of $n$, which specifies the system size $L = 2^n_{~}$, since the dimension of each index of $W^{(n)}_{~}$ is $2^L_{~}$. 
Thus the numerical storage of the order of $2^{4L}_{~}$ is required if one keeps $W^{(n)}_{~}$ faithfully. 
To overcome this limitation, the HOSVD is introduced in the formulation of the HOTRG, and each tensor is {\it compressed} to that of smaller dimensions~\cite{Xie_2012_HOTRG}. 

\begin{figure}[Htb]
  \centering
  \resizebox{4.3cm}{!}{\includegraphics{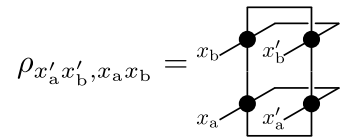}}
  \caption{ Graphical representation of the reduced density matrix in Eq.~(\ref{dml}). }
  \label{fig3}
\end{figure}

Suppose that we have reached a maximum $n$, where we can generate $M^{(n+1,n)}_{~}$ but cannot store the all tensor elements of $W^{(n+1)}_{~}$. 
Following the standard procedure in HOSVD proposed in Ref.~[\onlinecite{Xie_2012_HOTRG}], let us introduce a kind of density matrix
\begin{equation}
\rho^{(n+1)}_{x'_{\rm a} x'_{\rm b}, x^{~}_{\rm a} x^{~}_{\rm b} } = 
\sum_{x''_{\rm a} x''_{\rm b}   y^{~}_{\rm a} y^{~}_{\rm b} } 
M^{(n+1,n)}_{x^{~}_{\rm a} x^{~}_{\rm b} x''_{\rm a} x''_{\rm b} y^{~}_{a} y^{~}_{\rm b}} 
M^{(n+1,n)}_{x'_{\rm a} x'_{\rm b} x''_{\rm a} x''_{\rm b} y^{~}_{\rm a} y^{~}_{\rm b}} \, .
\label{dml}
\end{equation}
%
Figure \ref{fig3} exhibits the schematic picture of $\rho^{(n+1)_{~}}$; sewing three edges of two $M^{(n+1,n)}_{~}$, we obtain the {\it reduced} density matrix with respect to the pair of indices 
$\{ x_{\rm a}^{~} x_{\rm b}^{~} \}$ and $\{ x'_{\rm a} x'_{\rm b} \}$.
The RG transformation matrix is then constructed from the diagonalization 
\begin{equation}
\rho^{(n+1)}_{~} = U^{(n+1)}_{~} \, \Omega \, U^{(n+1)\dagger}_{~} \, ,
\label{rdm}
\end{equation}
where the eigenvalue matrix $\Omega$--- the entanglement spectrum in HOTRG--- 
is positive definite, and where $U^{(n+1)}_{~}$ is the corresponding orthogonal matrix~\cite{ent_spec}.
Let us assume that the diagonal elements $\Omega_\mu^{~}$ of $\Omega$ are aligned in the decreasing order. 
Normally, the decay in $\Omega_\mu^{~}$ with respect to $\mu$ is rapid enough, and it is possible to discard tiny eigenvalues $\Omega_\mu^{~} \ll 1$. 
We thus retain $D$ numbers of relevant eigenvalues in accordance with the standard DMRG scheme.

After the restriction $\mu \le D$, the orthogonal matrix 
$U^{(n+1)}_{x^{~}_{\rm a} x^{~}_{\rm b}, \, \mu}$ 
can be regarded as a RG transformation. 
It is naturally applied to $M^{(n+1,n)}_{~}$ in the manner
\begin{equation}
M^{(n+1,n)}_{\mu \mu'  y^{~}_{\rm a}  y'_{\rm a}}  = 
\!\!\!\!\!\!
\sum_{x^{~}_{\rm a} x^{~}_{\rm b} x'_{\rm a} x'_{\rm b}}
\!\!\!\!
[ U^{(n+1) \dagger}_{~} ]_{\mu, \, x^{~}_{\rm a} x^{~}_{\rm b}} 
M^{(n+1,n)}_{x^{~}_{\rm a} x^{~}_{\rm b} x'_{\rm a} x'_{\rm b} y^{~}_{\rm a} y'_{\rm a}} 
U^{(n+1)}_{x'_{\rm a} x'_{\rm b}, \,\mu'_{~}} \, ,
\label{cg}
\end{equation}
where the symmetry of the local vertex is assumed. 
We then obtain the renormalized vertex tensor 
$M^{(n+1,n)}_{\mu \mu'  y^{~}_{\rm a}  y'_{\rm a}}$, 
which has two renormalized indices $\mu$ and $\mu'$. 
Using $M^{(n+1,n)}_{~}$, we can join two of them horizontally, as we have done in Eq.~(1), to obtain $W^{(n+1)}_{~}$, which is an extended tensor whose linear dimension is $2L = 2^{n+1}_{~}$. 
Creating the density matrix for the vertical links and performing the RG transformation again, we obtain a renormalized vertex tensor $W^{(n+1)}_{\mu \mu'_{~} \nu \nu'_{~}}$. 
Such a process of system extension and RG transformation can be repeated for arbitrary times, within a practical computational time. 
It is expected that, as $n$ increases, the renormalized tensor $W^{(n)}_{~}$ approaches to that of the thermodynamic limit, $W^*_{~}$.

Throughout this article we use alphabetical indices, such as $x_{\rm a}^{~}$ and $y_{\rm b}^{~}$ for row/column of the original link variables, and Greek letters, such as $\mu$ and $\nu$, for the renormalized link variable whose degrees of freedom is $D$ at most. 
We often abbreviate indices, such as $W^{(n)}_{~}$, if the distinction between renormalized and unrenormalized tensors is apparent. 

In addition, we note that the partition function per site can be obtained in terms of normalization coefficients of renormalized vertex tensor $W^{(n)}_{~}$.
In order to avoid an explosion of the overall normalization of $W^{(n)}_{~}$, it is useful to impose the normalization
\begin{equation}
\sum_{\mu \nu} W^{(n)}_{\mu \mu \nu \nu}
= \sum_{\mu \nu} M^{ (n,n+1) }_{\mu \mu  \nu \nu}
=  1 .
\label{vnorm}
\end{equation}
for every iteration step.
Practically, we calculate a normalization coefficient, $\tilde{\gamma}_n = \sum_{\mu \nu} W^{ (n) }_{\mu \mu  \nu \nu}$ and then replace $W^{ (n) }_{\mu \mu'_{~} \nu \nu'_{~}} / \tilde{\gamma}_n \rightarrow W^{(n)}_{\mu \mu'_{~} \nu \nu'_{~}}$.
Also, we define $\gamma_{n+1} = \sum_{\mu \nu} M^{(n+1,n)}_{\mu \mu  \nu \nu}$ and  replace $M^{(n+1,n)}_{\mu \mu'_{~} \nu \nu'_{~}} / \gamma_n\to M^{(n+1,n)}_{\mu \mu'_{~} \nu \nu'_{~}}$.
For a given iteration number $n$, then, the partition function for the $L \times L$($L = 2^n_{~}$) system can be expressed as
$
Z(L, D) = \prod_{i=1}^{n} \gamma_i^{L^2/2^{(2i-1)}} \tilde{\gamma}_i^{L^2/2^{2i}}$,
where  traces of the unnormalized vertex are taken in both of the vertical and horizontal directions, respectively.
Note that the geometry of this partition function corresponds to a torus.
We can thus calculate the logarithm of the partition function per site as
\begin{equation}
\log z( L, D ) = \frac{\log Z(L, D)}{L^2} = \sum_{i = 1}^{n} \frac{1}{2^{2i-1}}(\log \gamma_i^{~} + \frac{1}{2}\log \tilde{\gamma}_i^{~} ).
\label{z_sum}
\end{equation}
The accuracy of the approximation is determined by the truncation error $\sum_{\mu > D} \Omega_{\mu}$.

%

\section{fixed point and doubling of the entanglement spectrum}
\label{fp}

Let us consider the fixed point of the HOTRG method in the off-critical regime, where the correlation length of the system is finite. 
Recall that the spatial width of the renormalized vertex tensor $W^{(n)}_{~}$ increases exponentially with respect to the number of extension $n$. 
This implies that $L$ exceeds the correlation length of the system $\xi$ above a certain number of iterations $n$, and finally $L \gg \xi$ is satisfied at the fixed point. 
Then the link variables corresponding to the two parallel edges of $W^{(n)}_{~}$ 
are spatially separated away beyond $\xi$, where the entanglement between them is negligible.
Similarly, it can be expected that the link variables around different corners become disentangled with each other. 
In this sense, the renormalized vertex tensor at the fixed point $W^*_{~}$ can be decoupled into four patches. 
When such a decoupling scheme is realized, the vertex tensor is mentioned as the corner double-line (CDL) tensor~\cite{ Levin_2007_TRG,Gu_2009_TEFR}. 

\begin{figure}[Htb]
  \centering
  \resizebox{4.4cm}{!}{\includegraphics{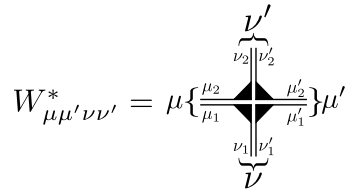}}
  \caption{ Double line representation of the renormalized vertex tensor. 
A solid triangle at a corner indicates a CTM and the link variable $\mu$ is represented by a combination of the two sub-link variables $\mu_1^{~}$ and $\mu_2^{~}$. }
  \label{fig4}
\end{figure}
\begin{figure}[Htb]
  \centering
  \resizebox{6.3cm}{!}{\includegraphics{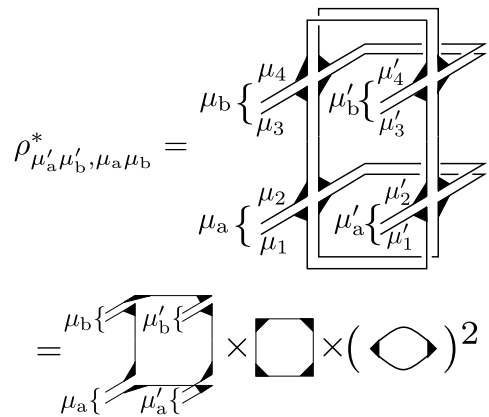}}
  \caption{ Double line representation of the reduced density matrix $\rho^*_{~}$ in 
Eq.~(\ref{dmctm}). The closed loops indicate trace of the products of CTMs, which give 
scalar constants.}
  \label{fig5}
\end{figure}

 For the unnormalized fixed-point vertex tensor $W^*_{~}$, there is a remaining entanglement in the two adjacent edges around each corner, because the distance between them is independent of the system size $L$. 
An essential point is that this entanglement around each corner is the same as that of the Baxter's CTM. 
Thus it is expected that the unnormalized fixed-point vertex tensor can be decomposed as
\begin{equation}
W^*_{\mu \mu' \nu \nu'} = 
\kappa \, 
C_{\mu^{~}_1 \nu^{~}_1}^{~} 
C_{\mu^{~}_2 \nu^{~}_2}^{~} 
C_{\mu'_1 \nu'_1}^{~} 
C_{\mu'_2 \nu'_2}^{~} \, , 
\label{cdl}
\end{equation}
where $C_{ \mu_1^{~} \nu_1^{~} }^{~}$ is a normalized CTM having the sub-link variables $\mu_1^{~}$ and $\nu_1^{~}$ of the effective dimension  $\sqrt{D}$.
In addition,
$\mu \equiv \{ \mu_{\rm 1}^{~}  \mu_{\rm 2}^{~} \}$, 
$\nu \equiv \{ \nu_1^{~}  \nu_2^{~} \}$, etc.,
represent the double line indices.
Here, we have assumed the isotropic model, for which $C$ is the real symmetric matrix. 
Also we have used the normalization of $C$ so that $\Tr \, [ C^4_{~} ] =1$ is satisfied.
The coefficient $\kappa$ denotes a normalization factor associated with the partition function in the thermodynamic limit. (See Eq.~(\ref{tm_eq}).)
In Fig.~\ref{fig4}, we show the schematic diagram of $W^*_{~}$, where $C_{\mu_1^{~} \nu_1^{~}}$ is illustrated as a ``{\sf L}''-shape line with a small solid triangle connecting the sub-link variables $\mu_1^{~}$ and $\nu_1^{~}$. 
In addition, such an index as $\mu$ in the vertex tensor consists of the double line index of $\{ \mu_{\rm 1}^{~} \mu_{\rm 2}^{~} \}$.
We have made a qualitative explanation on the CDL picture for the fixed-point renormalized tensor $W^*_{~}$. 
It is, however, possible to extract the picture theoretically, by considering matrix product states (MPS) that surround an unrenormalized tensor $W^{(n)}_{x x' y y'}$. The details will be discussed elsewhere~\cite{Nishino_2013_boundary_mps}.

On the basis of the CDL picture of the vertex tensor,  we can also see the decoupling in the reduced density matrix. 
Substituting Eq.~(\ref{cdl}) to Eq.~(\ref{dml}), we obtain 
\begin{equation}
\rho^*_{\mu'_{\rm a}\mu'_{\rm b}, \, \mu_{\rm a} \mu_{\rm b}} =  \alpha \, 
[ C^2_{~} ]_{\mu_1^{~} \mu'_1} 
[ C^2_{~} ]_{\mu_2^{~} \mu_3^{~}} 
[ C^2_{~} ]_{\mu'_2 \mu'_3} 
[ C^2_{~} ]_{\mu'_4 \mu_4^{~}} \, ,
\label{dmctm}
\end{equation}
where $\alpha \equiv \kappa^4_{~} ( \Tr \, [ C^2_{~} ] )^2_{~}$, and the indices of $\rho^*_{~}$ are given by
$\mu_{\rm a} \equiv \{ \mu_1^{~} \mu_2^{~} \}$, 
$\mu_{\rm b} \equiv \{ \mu_3^{~} \mu_4^{~} \}$, 
$\mu'_{\rm a} \equiv \{ \mu'_1 \mu'_2 \}$, and
$\mu'_{\rm b} \equiv \{ \mu'_3 \mu'_4 \} $.
This equation can be easily understood by the graphical representation, which is depicted in Fig.~\ref{fig5}. 
A small solid triangle at the junction of two lines represents a CTM and the connected lines indicate contraction of the matrix indices. 
Note that a closed loop gives a trace of  matrix product of CTMs, which yields a scalar constant.
It should be noted that the normalization condition of Eq. (\ref{vnorm}) is equivalent to $\Tr \, [ C^4_{~} ] =1$. 
Once the CDL picture is well established, we always obtain $\gamma_n^{~} = \Tr \, [ C^4_{~} ]  = 1$. 
In other words, there is no correction term to the free energy,  after the decoupled fixed point is reached. 

Let us diagonalize the reduced density matrix with help of the CDL picture. 
As was discussed by Baxter\cite{book:Baxter_1982_CTM}, the CTM has a proper thermodynamic limit in the off-critical regime. 
Thus we can write
\begin{equation}
C = V \, \Lambda \, V^\dagger_{~} \, ,
\end{equation}
where $\Lambda $ is a diagonal eigenvalue matrix with the normalization $\Tr \, [ \Lambda^4_{~} ] = 1$ and $V$ is the corresponding orthogonal matrix. 
A key point is that  the matrix rank of $\rho^{*}_{~}$ effectively reduces from $D^2_{~}$ to $D$, because of the CDL property; 
in Eq.~(\ref{dmctm}), the rank of the matrix associated with the indices pairs $\{ \mu_2^{~} \mu_3^{~} \}$ and $\{ \mu'_2 \mu'_3 \}$ is just one and its eigenvalue is given by unity, which originates from $\Tr \, [ C^4_{~} ] =1 $. 
Thus the descendant matrix that we have to treat is $\alpha \, C^2_{~} \otimes C^2_{~}$ carrying the index pairs $\{ \mu_1^{~} \mu'_1 \}$ and $\{ \mu_4^{~} \mu'_4 \}$. 
We can now diagonalize $\rho^*_{~}$ by the unitary matrix $U^*_{~} = V \otimes V$;
\begin{equation}
\Omega^*_{~} = \alpha \, \Lambda^2_{~} \otimes \Lambda^2_{~} \, ,
\label{doubling}
\end{equation}
where $\Omega^*_{~}$ is the fixed point spectrum of Eq.~(\ref{rdm}). 
This implies that the entanglement spectrum in HOTRG is described  by the doubling of the CTM spectrum.

For a class of  integrable models in the  off-critical regime, the eigenvalue spectrum $\Lambda$ in the bulk limit  is exactly obtained as an infinite direct product of the $2 \times 2$ diagonal matrix as follow~\cite{book:Baxter_1982_CTM}
\begin{equation}
\Lambda = \bigotimes_{n = 1}^{\infty} \begin{pmatrix} 1 & 0 \\ 0 & q^{c_n^{~}}_{~} \end{pmatrix} \, , 
\label{spectrum}
\end{equation}
where $c_n^{~}$ is a model-dependent sequence. 
We have used the normalization such that the largest eigenvalue is unity.
The value of $q~( 0 < q < 1 )$ qualitatively represents a distance from the critical point and it is related to interaction parameters of the model. 
For the case of the Ising model~\cite{book:Baxter_1982_CTM,Tsang_1979_2DIsing}, the sequence $c_n$ is given by
\begin{equation}
c_n^{~} = \left\{ \begin{matrix} n & ( T < T_{\rm c} ) \\ 2n - 1 & ( T > T_{\rm c} ) \end{matrix} \right.~
\label{sequence}
\end{equation}
and $q$ is the nome of the elliptic function with the modulus 
$k = \sinh^{-2}(2/T)$.\cite{duality}
In the next section, we will demonstrate that the doubling of the spectrum actually occurs for Ising model by numerical computations.

\begin{figure}[Htb]
  \centering
  \resizebox{4.6cm}{!}{\includegraphics{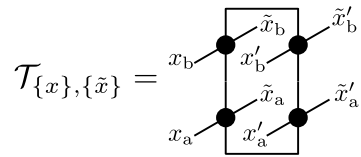}}
  \caption{ Graphical representation of the $L$-layer transfer matrix ${\cal T}$ in Eq.~(\ref{lltm}). }
  \label{fig6}
\end{figure}

In the remaining part of this section, we discuss the relation between the HOTRG formulation and the row-to-row (or column-to-column) transfer matrix under the periodic boundary condition. 
Let us introduce a single-layer row-to-row transfer matrix
\begin{equation}
\tau_{\{x\} \{x'\}_{~}}^{~} = \sum_{\{y\}} \prod_{i = 1}^{4L} W_{x_i^{~} x_i^{\prime} y_i^{~} y_{i+1}^{~}} 
\label{sltm}
\end{equation}
of length $4L$, where we have introduced notations $\{x\} = \{x_1^{~} \cdots x_{4L}^{~} \}$, 
$\{x'\} = \{x'_1 \cdots x'_{4L} \}$ and $\{y\} = \{y_1^{~} \cdots y_{4L}^{~} \}$. 
We impose the periodic boundary condition $y_{4L+1}^{~} = y_1^{~}$, and the configuration sum is taken over for the horizontal links $\{y\}$.
The thermodynamic property of the cylindrical system is described by the maximum eigenvalue $\lambda_{\rm max}^{~}$ of $\tau$ and the corresponding eigenvector $\vec{v}$.

In order to see the relation between the transfer matrix $\tau$ and the reduced density matrix in Eq.~(\ref{dml}), we introduce the $L$-layer transfer matrix 
\begin{equation}
{\cal T} \equiv \tau^L_{~}.
\label{lltm}
\end{equation}
As shown in Fig.~\ref{fig6}, we can represent  ${\cal T}^{(n)}$ as a contraction of the four unrenormalized $W^{(n)}_{~}$, or equivalently a contraction of two $M^{(n+1,n)}_{~}$, where the link variables are assigned as 
$x_{\rm a} = \{x_1^{~} \cdots x_L^{~} \}$, 
$x_{\rm b} = \{x_{L+1}^{~} \cdots x_{2L}^{~} \}$, 
$x'_{\rm b} = \{x_{2L+1}^{~} \cdots x_{3L}^{~} \}$, and 
$x'_{\rm a} = \{x_{3L+1}^{~} \cdots x_{4L}^{~} \}$.
An essential point is that,  for $L \gg \xi$,  the largest eigenvalue $\lambda_{\rm max}^{~}$ of $\tau$ becomes dominant in ${\cal T}$. 
Thus, in the thermodynamic limit, we have ${\cal T} \sim (\lambda_{\rm max}^{~})^L \vec{v}  \vec{v}^\dagger $, where the link variables of the row and column indices of ${\cal T}$ are disentangled with each other. 
For a sufficiently large system size $L$, the matrix rank of ${\cal T}$ collapse to one, which is another aspect of the CDL property. 
The CDL property of ${\cal T}^*$ is basically maintained through the RG transformation $x_{\rm a}^{~} x_{\rm b}^{~} x'_{\rm a} x'_{\rm b}$ $\rightarrow$ $\mu_{\rm a}^{~} \mu_{\rm b}^{~} \mu'_{\rm a} \mu'_{\rm b}$.
This suggests that the CDL property of the renormalized vertex tensor $W^*_{~}$ at the fixed point can be quantitatively evaluated by solving the eigenvalue problem of ${\cal T}^*_{~}$. 

In the representation of the renormalized indices $\{\mu\}$,  we further rewrite the reduced density matrix 
\begin{eqnarray} 
\rho^{* }_{\mu^{~}_{\rm a} \mu^{~}_{\rm b}, \, \mu'_{\rm a} \mu'_{\rm b}} 
= \sum_{\tilde{\mu}_{\rm a} \tilde{\mu}_{\rm b}} 
\left[ {\cal T}^*_{~} \right]_{\mu^{~}_{\rm a} \mu^{~}_{\rm b} \mu'_{\rm a} \mu'_{\rm b}, \, 
\tilde{\mu}_{\rm a} \tilde{\mu}_{\rm b} \tilde{\mu}_{\rm a} \tilde{\mu}_{\rm b}} \, , 
\label{dm_tm}
\end{eqnarray} 
where the summation corresponds to sewing of the bottom side of ${\cal T}^*$.
Then, with the help of the CDL representation of ${\cal T}^*_{~}$ and 
$\rho^*_{~}$, we can show the relation 
\begin{eqnarray}
{\cal T}^*_{~} \rho^*_{~} = \kappa^{4}_{~} \rho^*_{~} \, ,
\label{tm_eq}
\end{eqnarray}
where all of the closed loops give contribution of $\Tr \, [ C^4_{~} ] =1$.
Thus, the reduced density matrix itself is the eigenvector of the renormalized transfer matrix.
Since $\tau$ includes $4L$ number of the local vertices, we can also verify that the 
relation around the eigenvalue 
$\kappa^4_{~} = (\lambda_{\rm max}^{~})^L = z^{4L^2_{~}}_{~}$,
where $z$ denotes the partition function per site.

\section{Numerical results in off critical region}
\label{des}

In order to confirm the CDL picture in HOTRG, we deal with the spatially uniform Ising model on the square lattice, where we represent the model as a symmetric $2$-state vertex model. 
The local vertex weight $W_{x \, x' y \, y'}^{~}$ is given by 
\begin{equation}
W_{x \, x' y \, y'}^{~} = \sum_{ \alpha = 1 }^2 
g_{\alpha x}^{~} \, 
g_{\alpha x'}^{~} \, 
g_{\alpha y}^{~} \, 
g_{\alpha y'}^{~} \, , 
\end{equation}
where $g$ is a $2 \times 2$ matrix 
\begin{equation}
g = \begin{pmatrix}
  \sqrt{\cosh (1/T)} & \sqrt{\sinh (1/T)}  \\ 
  \sqrt{\cosh (1/T)} & -\sqrt{\sinh (1/T)} 
\end{pmatrix} \, ,
\end{equation}
which is dependent on the temperature $T$.
Note that the critical temperature of this model is $T_{\rm c}^{~} = 2/\ln(1+\sqrt{2})$.~\cite{Onsager_1944_2DIsing}

\begin{figure}[Htb]
  \centering
  \resizebox{8cm}{!}{\includegraphics{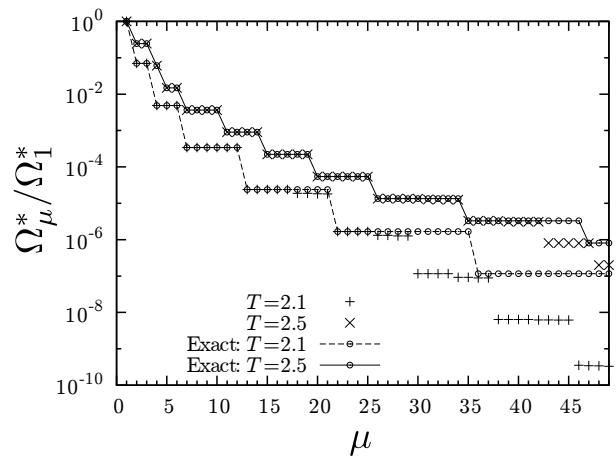}}
  \caption{Entanglement spectrum $\Omega^*_\mu $ of the reduced density matrix $\rho^*_{~}$ for $D = 49$. 
The open circles with the solid and broken lines respectively show the doubling spectra of the corresponding CTMs. 
}
\label{spectrum_D49}
\end{figure}

As in Eq.~(\ref{doubling}), the analytic form of the entanglement spectrum at the fixed point in the off critical region is given by the doubling of the CTM spectrum~(\ref{spectrum}). 
According to the degeneracy structure in the CTM spectrum, the doubled spectrum $\Omega^*_\mu$ has the following sequence of the degeneracy
\begin{eqnarray*}
&&{\tt 1 \quad  2 \quad 1 \quad 2  \quad 4  \quad ~ \, 4 \quad 5 \, \cdots } \qquad T > T_{\rm c}^{~}\\
&&{\tt 1 \quad  2 \quad 3 \quad 6  \quad 9  \quad 14 \quad ~~ \cdots } \qquad T < T_{\rm c}^{~} \, .
\end{eqnarray*}
For two typical temperatures $T = 2.1$ ($< T_{\rm c}^{~}$) and $T = 2.5$ ($> T_{\rm c}^{~}$),  we plot the above analytic sequence in Fig.~\ref{spectrum_D49} as small circles with the broken line ($T = 2.1$) and with solid line ($T = 2.5$). 
In the figure, we have used the scale of the vertical axis such that the largest spectrum, $\Omega_1^{*}$, is unity.

For $T = 2.1$ and $T = 2.5$, we have performed HOTRG computations up to $n = 50$  ($L\sim10^{15}$) 
with the number of the retained basis $D = 49$, which is sufficient for precise estimation of thermodynamic quantities.
Details about convergence of HOTRG iterations are presented in Appendix A.
Here, we just note that  $n= 50$ is sufficient to obtain the fixed-point tensors, except for irrelevant gauge degrees of freedom associated with the degenerating eigenvalues. 

In Fig.~\ref{spectrum_D49}, the plus and cross symbols represent $\Omega^*_\mu$ obtained by HOTRG\cite{symmetric}, which shows a good agreement with the analytic one up to $\mu \sim 30$. 
This agreement is a numerical evidence of the CDL picture in the off critical region. 
The deviation from the exact result in the large $\mu$ region is attributed to the perturbation due to the cut off $D = 49$. 

To further confirm the CDL picture, we also evaluated the rank of the $L$-layer transfer matrix ${\cal T}$.  
As was discussed in the previous section, the rank should be reduced into one, when the number of iteration $n$ exceeds a certain number associated with the correlation length. 
We numerically observed the spectrum of ${\cal T}$ for $D = 10$, where the dimension of ${\cal T}$ is proportional to $\mathcal{O}( D^{4}_{~} )$. 
We have verified that the all eigenvalues of ${\cal T}$  except for the maximum one collapses to zero after ${\cal T}$ converges to ${\cal T}^*_{~}$ (the numerical result is not presented here).
This result supports the CDL picture of the vertex tensors at the fixed point.

Here, we would like to comment on the relevance to the 1D quantum system with a gapful groundstate. 
Since there is well-established correspondence between 1D quantum systems and 2D classical systems, one can expect that the doubling of the entanglement spectrum for the 1D quantum system under the periodic boundary condition.
In particular, it should be remarked that, for the integrable model,  the eigenvector of the Hamiltonian and the corresponding transfer matrix are equivalent. 
We have actually formulated an HOTRG-like tensor RG,  detail of which is presented in Appendix \ref{1d_quantum}, and performed a numerical computation for the 1D transverse-field Ising model in the off-critical region.
We then confirmed that the corresponding entanglement spectrum of the 1D transverse-field Ising model is equivalent to Fig. \ref{spectrum_D49}.

\section{Critical Region}
\label{fds}

In the critical region $T \sim T_{\rm c}^{~}$  the correlation length diverges as $\xi \sim | T-T_{\rm c}^{~} |^{-\nu}$. 
Therefore the coupling between the CTMs is non-negligible at $T_{\rm c}^{~}$ regardless of the size $L = 2^n_{~}$, and thus the CDL decoupling picture for the renormalized tensors would not be appropriate any more. 
The numerical data calculated at  $T_{\rm c}^{~}$, however, shows that the spectrum $\Omega_\mu^{~}$ and the vertex tensors actually converge within $n = 50$ iterations with the decoupling of CDL as shown in Fig. \ref{rank}. 
This is because the cutoff $D$ introduces an effective length scale $\xi_{\rm eff}^{~}$ into the system. 
While the system size  $L = 2^n_{~}$ is less than $\xi_{\rm eff}^{~}$, the finite size scaling behavior may be observed.
After $L$ exceeds  $\xi_{\rm eff}^{~}$, a quasi-off-critical behavior emerges in the thermodynamic quantities.

\begin{figure}[Htb]
  \centering
  \resizebox{8cm}{!}{\includegraphics{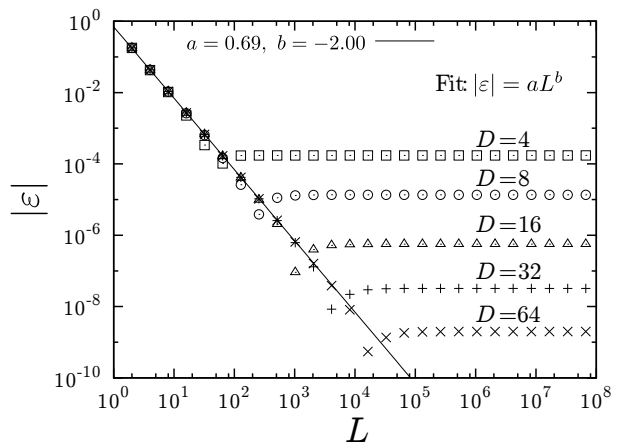}}
  \caption{  Absolute value of relative error $\varepsilon( L, D )$ in Eq.~(\ref{lpf}) at $T_{\rm c}^{~}$. 
The solid line shows the fitting to $| \varepsilon | \sim a \, L^b_{~}$, where the best fit is obtained for $a = 0.69$ and $b = - 2.00$.
}
\label{finite_L_region_D_region}
\end{figure}

We first analyze the above crossover in the free energy level. 
Remember the partition function per site $z( L, D )$ in Eq.~(\ref{z_sum}), which can be calculated by normalization constants in HOTRG iterations. 
Because of the finite size effect and the presence of the cut-off $D$, $z( L, D )$  at $T_{\rm c}^{~}$ contains some deviation from the exact partition function per site $z_{\rm ex}^{~}$ in the thermodynamic limit~\cite{Onsager_1944_2DIsing}. 
We observe the relative error 
\begin{equation}
\varepsilon( L, D ) \equiv 1-\frac{\log z( L, D )}{\log z_{\rm ex}^{~}} \, ,
\label{lpf}
\end{equation}
where $\log z( L, D )$ is equivalent to the free energy per site except for the overall sign and temperature factors. 
Figure \ref{finite_L_region_D_region} shows the $L$-dependence of $| \varepsilon( L, D ) |$ for $D = 4 \sim 64$. 
A clear crossover can be seen from the $L$-dependent (or small $L$) 
to the $D$-dependent (or large $L$) region where $\varepsilon( L, D )$ converges to a constant value with respect to $L$.
For the small $L$ region, we have the fitting result of $| \varepsilon | \sim a \, L^b_{~}$ with $a = 0.69$ and $b = - 2.00$,   which is consistent with the standard finite-size-scaling behavior $|\varepsilon(L,\infty)| \sim L^{-2}$. 

\begin{figure}[Htb]
  \centering
  \resizebox{8cm}{!}{\includegraphics{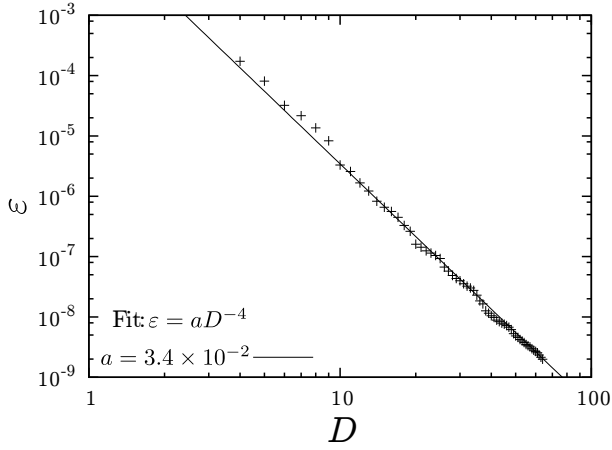}}
  \caption{ Relative error $\varepsilon( L, D )$ in Eq.~(\ref{lpf})  at $T_{\rm c}^{~}$ in the large-$L$ limit. 
The solid line shows a fitting line $a D^{-4}_{~}$ with $a = 0.034$. 
}
\label{re_err_finite_D_tc}
\end{figure}

We also analyze the $D$-dependence in the sufficiently large $L$ region, as shown in Fig.~\ref{re_err_finite_D_tc}. 
The error is well fitted by the function $\varepsilon(\infty,D)  = a D^{-4}_{~}$, where the prefactor is calculated as $a = 0.034$. 
In order to capture the background of this functional form, we assume the $D$-dependence
\begin{equation}
\xi_{\rm eff} \sim D^{\theta} 
\label{xidscal}
\end{equation}
for the effective length scale $\xi_{\rm eff}^{~}$ in the large $L$ fixed point. 
Since the crossover occurs around $L \sim \xi_{\rm eff}$, we have the relation $\varepsilon( \xi_{\rm eff}, D ) \sim \xi^{-2}_{\rm eff} \sim D^{-2\theta}_{~}$, which  specifies the $D$-dependence of the error for  $L\gg \xi_{\rm eff}$. 
In comparison with the fitting result in Fig.~\ref{re_err_finite_D_tc}, we can read off $\theta = 2$, which is consistent with the finite-$\chi$ scaling based on the MPS variation for the 1D transverse-field Ising model~\cite{Tagliacozzo_2008_EE_MPS,Pollmann_2009_FES,Pirvu_2012_FSS-FES}.

\section{ Entanglement entropy at criticality }
\label{eec}

\begin{figure}[Htb]
  \centering
  \resizebox{8cm}{!}{\includegraphics{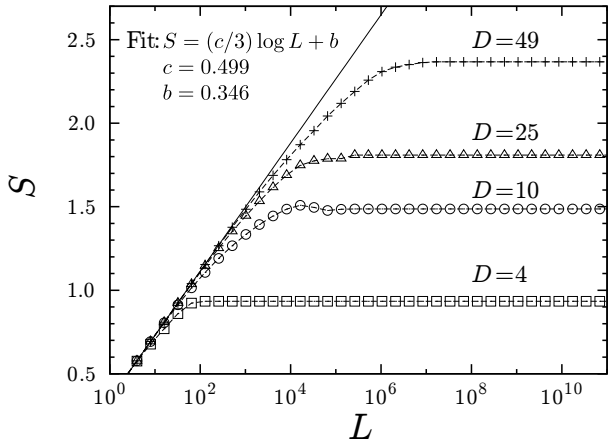}}
  \caption{ Entanglement entropy in $S( L, D )$ at $T_{\rm c}^{~}$. 
The solid line shows the linear fit to the function $(c/3) \log L + b$. 
}
\label{ent_ent_tc_finite_L}
\end{figure}

We next observe the entanglement entropy at criticality~\cite{Holzhey_1994_EE_CFT,Osterloh_2002_EE,Vidal_2003_EE,Calabrese_2004_EE_CFT,Refael_2004_EE_CFT,Amico_2008_EE_refiew}. 
As  was shown in Eq.~(\ref{tm_eq}), the reduced density matrix $\rho^*$ at the fixed point is the eigenvector of the row-to-row transfer matrix under the periodic boundary condition. 
In this sense,  $\rho^{(n)}$  well approximates the eigenvector of ${\cal T}^{(n)}$ even at the critical point.
We therefore define an entanglement entropy as
\begin{equation}
S( L, D ) = -\sum_\mu^{~} {\Omega}_\mu^2 \, \log \, {\Omega}^2_\mu \, ,
\label{ent_ent}
\end{equation}
where  $\Omega_\mu^{~}$ are the eigenvalues of the reduced density matrix for $L=2^n$, and are normalized so that $\sum_{\mu=1}^{D} \Omega_\mu^{2} = 1$.  
Note that the linear dimension of the transfer matrix is $4L$, and therefore $S( L, D )$ corresponds to the bipartition of $2L+2L$.
If $D$ is sufficiently large, i.e. $D = \infty$, we expect that the leading term of $S( L, \infty )$ follows the CFT prediction\cite{Calabrese_2004_EE_CFT} 
\begin{equation}
S_{\rm CFT}^{~}( 2L )  \sim \frac{c}{3} \log ( 2L / a ) \, ,
\label{eefss}
\end{equation}
where $c$ is the central charge and $a$ is a microscopic cut-off scale.
Figure \ref{ent_ent_tc_finite_L} shows  $S( L, D )$ calculated at $T_{\rm c}^{~}$.
In the region $L < \xi_{\rm eff}^{~}$,  a clear $\log L $ dependence is observed  in $S( L, D )$. 
A linear fitting for the case of $D = 49$ in the window $4 \le L \le 32$ yields $c = 0.499$,  which is consistent with the central charge $c = 1/2$ of the Ising universality.

\begin{figure}[Htb]
  \centering
  \resizebox{8cm}{!}{\includegraphics{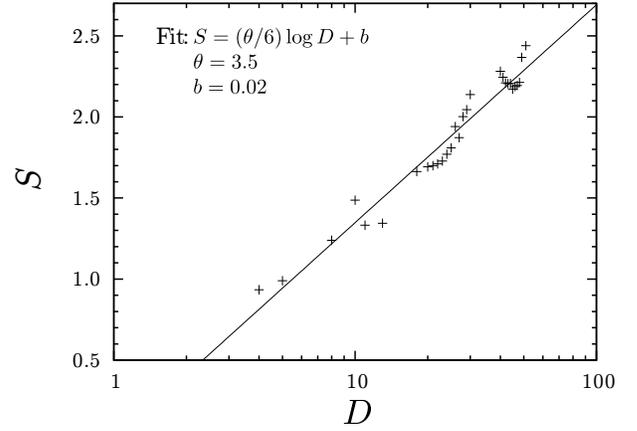}}
  \caption{ The $D$-dependence of $S( \infty, D )$ at $T_{\rm c}^{~}$. The line represent the least square fitting to the function $( \theta/6 ) {\rm log}D + b$.
}
\label{ent_ent_tc}
\end{figure}

Similarly to the free energy, let us observe the $D$-dependence of the entanglement entropy $S( \infty, D )$ for the sufficiently large $L$.
According to CFT~\cite{Calabrese_2004_EE_CFT}, the entanglement entropy for a single strip with two boundary points in the vicinity of $T_c$ is given by
\begin{equation}
S_{\rm CFT}^{~}( \xi )  \sim  \frac{c}{3} \log \xi \, ,
\label{eexicft}
\end{equation}
where the system size and the length of the strip are assumed to be much longer than the correlation length of the system.
Substituting  Eq. (\ref{xidscal}) into Eq. (\ref{eexicft}), we obtain the finite-$D$ scaling of the entanglement entropy  for $L\gg \xi_{\rm eff}$ as
\begin{equation}
S( \xi_{\rm eff}^{~}, D )  \sim  \frac{c}{3}  \, \theta \, \log D .
\label{eedscal}
\end{equation}
Figure~\ref{ent_ent_tc} shows the  $D$-dependence of the entanglement entropy $S( \infty, D )$ calculated by HOTRG.
Although the plotted data are rather scattered, the overall behavior is consistent with the function $(\theta/6)  {\rm log}D + b$, with which we have the fitting result of $\theta \simeq 3.5$ and $b\simeq 0.02$.
However, it should be remarked that the estimated exponent,  $\theta \simeq 3.5$,  is {\it not consistent} with $\theta = 2$ obtained from  $\varepsilon( L, D )$ for the free energy.

A reason for this inconsistency would be attributed to the crossover of the reduced density matrix around $L \sim \xi_{\rm eff}$. 
As was discussed in Sec.~II, the accuracy of the free energy is determined by HOTRG iterations up to $L \sim \xi_{\rm eff}^{~}$, where the reduced density matrix in Eq.~(\ref{rdm}) has the full dimension of $D^2$; an example of this region is shown in $L \lesssim 10^{7}$ of Fig. \ref{rank} in Appendix A. 
Thus, $\xi_{\rm eff}^{~}$ is evaluated by the reduced density matrix of dimension $D^2$, and thus the proper finite-size-scaling result with $\theta = 2$ is observed. 
On the other hand,  the effective matrix dimension of the reduced density matrix for $L > \xi_{\rm eff}^{~}$ collapses to $D$ even at $T_{\rm c}^{~}$ as shown in Fig. \ref{rank}, where the CDL decoupling  effectively occurs.
One may afraid that this reduction of the effective dimension might possibly induces a redunction of the entanglement entropy in $L > \xi_{\rm eff}^{~}$.
As is shown in Fig.~\ref{ent_ent_tc_finite_L}, however,  $S( L, D )$ is approximately non-decreasing function of $L$, and is saturated toward a fixed point value.
This suggests that,  at the critical point,  the crossover behavior around $L \sim  \xi_{\rm eff}^{~}$ is responsible for a decoupling of the tensor $W^{(n)}_{~}$ to CTMs containing a nontrivial effective length scale. 

\begin{figure}[hbt]
  \centering
  \resizebox{8cm}{!}{\includegraphics{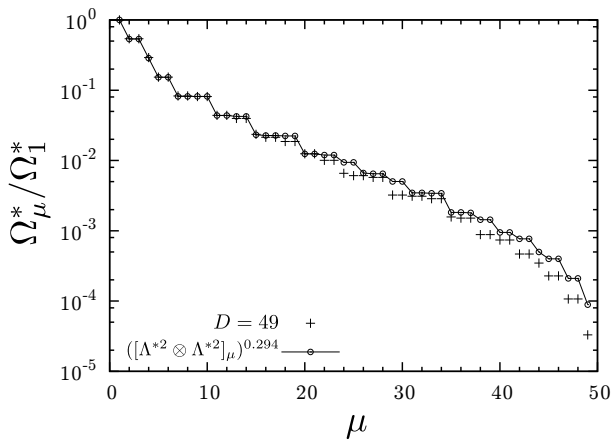}}
  \caption{The entanglement spectrum  $\Omega^{*}_{~}$ with $D = 49$ at $T_{\rm c}^{~}$.
For comparison, a nontrivial power of the doubling of the CTM spectrum $\Lambda^*$ obtained by a CTMRG calculation under the condition $m = 7$ is presented as a solid line with circles.
}
\label{appndxb1}
\end{figure}

In order to analyze the CDL property at $T_{\rm c}^{~}$ in detail, we present the entanglement spectrum with  $D = 49$ after the numerical convergence, in Fig.~\ref{appndxb1}. 
We have also performed a corner transfer matrix renormalization group~\cite{Nishino_1996_CTMRG,Nishino_1996_CTMRG_scaling} (CTMRG) calculation at $T_{\rm c}^{~}$ with $m = 7$, where $m$ is the number of the retained bases in the CTMRG calculation,  and obtain the eigenvalues $\Lambda_\mu^{*}$ of the CTM.  
An interesting point in Fig.~\ref{appndxb1} is that the entanglement spectrum $\Omega_\mu^{*}$(plus symbols) and the doubling spectrum of  $\Lambda_\mu^{*}$(circles) exhibit the correspondence
\begin{equation}
 \Omega_\mu^{*} \simeq \left( \, [ \Lambda^{*2} \otimes \Lambda^{*2} ]_{\mu}  \, \right)^{0.294}_{~} \, ,
\label{esattc}
\end{equation}
with the nontrivial power.
This fact shows that the entanglement spectrum of HOTRG at $T_{\rm c}^{~}$ in the region $L \gg \xi_{\rm eff}^{~}$  deviates from a naive expectation, $ \Omega_\mu^{*}  \simeq \Lambda^{*2} \otimes \Lambda^{*2}$, although the CDL picture  holds for $L \gg \xi_{\rm eff}^{~}$ even at $T_{\rm c}^{~}$.
In particular, the power 0.294 indicates that the HTORG spectrum  maintains more entanglement than the CTMRG with $m=7$.

As is in the MPS variational formulation of the 1D transverse field Ising model~\cite{Tagliacozzo_2008_EE_MPS}, we have confirmed that the correlation length at the fixed point of CTMRG  scales as $\xi_{\rm CTM}^{~} \sim m^2_{~}$, where $m$ is the number of block spin state kept in the CTMRG\cite{Nishino_1996_CTMRG_scaling}. 
Thus, we may expect that the doubling of the CTM spectrum in HOTRG would draw the length scale $\xi_{\rm CTM}^{~} \sim ( \sqrt{D} )^2_{~} \sim D$. 
However, the numerical result of Eq. (\ref{esattc}) indicates that the effective length scale at the  HOTRG fixed point has much longer length scale than the naive expectation $\xi_{\rm CTM}^{~}$.
Indeed, the entanglement entropy  holds the value acquired in the region $L < \xi_{\rm eff}^{~}$, even after $L \gg \xi_{\rm eff}^{~}$.
Thus it is concluded that the fixed point of HOTRG retains the scale of order of $\xi_{\rm eff}^{~} > \xi_{\rm CTM}^{~}$, although the fixed-point vertex weight collapses to the tensor well described by the CDL picture.
At the present stage, however, it is difficult to clarify the detailed mechanism of such nontrivial CDL behavior of the vertex tensor.

\section{Conclusions and discussions}

We have studied the structure of the entanglement spectrum in HOTRG for  the 2D classical vertex model. 
In the off-critical region, we have clarified that the spectrum at the fixed point, where the renormalized tensor converges, is described by the doubling of the CTM spectrum having the effective dimension $\sqrt{D}$. 
This is in accordance with the CDL decoupling picture in the renormalized vertex tensor, as was discussed in the tensor RGs~\cite{Levin_2007_TRG, Gu_2009_TEFR}. 
The reduction of the matrix ranks in the reduced density matrix and the transfer matrix  also confirms the CDL picture. 
Moreover, the same doubling of the entanglement spectrum is verified for the  groundstate of the 1D transverse field Ising model in the off-critical region.

We have also investigated the finite-$D$ scaling at the criticality, where the cutoff $D$ introduces an effective length scale $\xi_{\rm eff}^{~}$.
For the 2D Ising model, we confirmed  $\xi_{\rm eff}^{~} \sim D^{\theta}_{~}$ with  $\theta = 2$ in the free energy level, where the exponent $\theta=2$  is determined within the range $L < \xi_{\rm eff}^{~}$.
Also,  $\theta=2$ is consistent with the finite-$\chi$ scaling of the MPS variational method for the 1D quantum Ising model at the critical point~\cite{Tagliacozzo_2008_EE_MPS}. 
On the other hand, the finite-$D$ scaling applied to the entanglement entropy suggests a nontrivial exponent $\theta\sim 3.5$, although the numerical result shows the CDL decoupling of the vertex tensors.
A reason for the discrepancy of $\theta$ is attributed to the fact that the entanglement spectrum is described by the  doubling of the CTM spectrum with the nontrivial power, as in Eq. (\ref{esattc}).
Then, a key point is that, at the critical point, geometry of the reduced density matrix may affect the structure of the effective fixed point of HOTRG, because the HOTRG algorithm accumulates deviations originating from the geometry during iterations.
For example, the geometry of the wavefunction in the MPS variation for the 1D quantum system\cite{Tagliacozzo_2008_EE_MPS,Pirvu_2012_FSS-FES} is the half-infinite cylinder, while that of Eq. (\ref{rdm}) is a finite size cluster where one end of the cylinder is bound off.
We think that this difference of the geometry is a possible reason for the nontrivial exponent $\theta = 3.5$ of HOTRG.
A similar geometrical effect is also expected for the HOTRG-like algorithm for the 1D quantum system, which was described in Appendix B, at the criticality.
Nevertheless, we would like to leave the detailed analysis as a future issue.

In this article, we have not considered the second renormalization group, which takes account of the entanglement between the vertex weight and surrounding environment. 
In contrast to HOTRG, the reduced density matrix in the second renormalization does not undergo the reduction of the matrix rank, which is a possible reason for the improvement of the accuracy in the second renormalization.
Also, we have not discussed HOTRG in higher dimensions.
We can expect that the renormalized tensor is described by  ``corner transfer tensor'' in the fixed point level.
However, the nature of the spectrum of the corner transfer tensor is not well-understand. 
For further analysis of HOSRG/TRG, it may be interesting issue to discuss the network structure of the tensors.

\begin{acknowledgments}
H. U. acknowledges discussions with I.~Maruyama.
T. N thanks G. Vidal for valuable discussion. 
This work was supported in part by Grants-in-Aid (Nos. 25800221, 23540442, 25400401, 24$\cdot$02815 and 23340109) from the Ministry of Education, Culture, Sports, Science and Technology of Japan. 
\end{acknowledgments}

\appendix

\section{Convergence of HOTRG iterations}

\begin{figure}[Htb]
  \centering
  \resizebox{8.6cm}{!}{\includegraphics{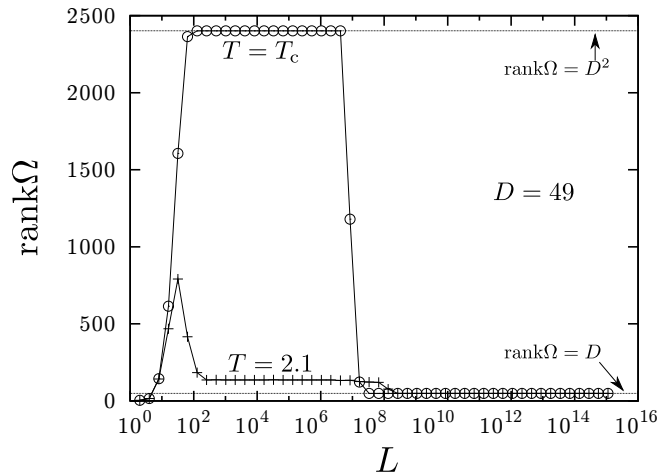}}
  \caption{ System size dependence of ${\rm rank}\Omega$ in Eq. (\ref{rdm}), where $\Omega_\mu/\Omega_1 < 10^{-15}$ is regarded as zero numerically. 
}
\label{rank}
\end{figure}

We present typical behaviors of the HOTRG iteration at $T=2.1$ and $T_c$.
As was discussed in Sec. II,  the matrix rank of the reduced density matrix (\ref{rdm}) is expected to crossover from $D^2$ to $D$, reflecting the CDL decoupling.
In Fig. \ref{rank}, thus, we plot the rank of $\Omega$ in Eq. (\ref{rdm}) as a function of the system length $L=2^n$, where the maximum dimension of the renormalized vertex tensor is  $D=49$.
In the figure, a numerical threshold for judging the zero eigenvalue is $\Omega_\mu/\Omega_1 < 1.0 \times 10^{-15}$.

In the region of $L \le 64$, the tensor dimension increases exponentially without any cutoff of the tensors. 
For $T=2.1(<T_c)$, the rank of $\Omega$ rapidly collapses, as $L$ increases beyond the correlation length of the system.
Indeed, the rank of  $\Omega$ becomes stable for $L>256$. 
However, it should be remark that this value of the rank is $2D$ rather than $D$.
This is because the all spectra below $T_c$ has the trivial double degeneracy associated with the $Z_2$ symmetry. 
As  $n$ further increases, the $Z_2$ symmetry of the spectrum is spontaneously broken around $L\sim 10^8$, triggered by the numerical error.
Then, the rank of $\Omega$ finally arrives at $D$, which indicates the proper CDL fixed point with the broken $Z_2$ symmetry. 
We thus confirmed the CDL decoupling at the off-critical region.

At $T=T_c$, the correlation length is intrinsically infinite. 
Accordingly, the rank of $\Omega$ maintains $D^2$ in the region of $ 64 \lesssim L \lesssim 10^7$.
As was mentioned in Sec. IV, however, a finite $D$ introduces an effective correlation length, and, as $L$ increases beyond it, the CDL decoupling may occur. 
In the region of $ L \gtrsim 10^8$, it can be actually seen that the rank of $\Omega$ drastically collapse to $D$.
Thus, we have also confirmed the decoupling scheme of Eq. (\ref{cdl}), even at the critical temperature. 

\section{HOTRG-like RG algorithm for 1D quantum systems}
\label{1d_quantum}
As was mentioned in Sec. IV, it is possible to formulate a 1D quantum system version of HOTRG algorithm.
Let us consider a 1D quantum spin model of length $4L$ described by a Hamiltonian having a nearest-neighbor interaction under the periodic boundary condition, where $L = 2^n$.
It is useful to introduce the matrix product operator (MPO) representation~\cite{Verstraete_2004_MPDO, McCulloch_2007_MPO}  of a Hamiltonian
\begin{equation}
\hat{\mathcal{H}} = \sum_{\{ \sigma \} \{ \sigma' \}}\Tr \Big[ \prod_{i=1}^{4L} \mathcal{W}_{\sigma_i \sigma'_i} \Big] \ketbra{\sigma}{\sigma'}, 
\end{equation}
where $\mathcal{W}_{\sigma_i \sigma'_i}$ is the MPO constructed from the local Hamiltonian at $i$ and $i+1$ sites, and $\{ \sigma \} \equiv \{ \sigma_{1}, \cdots, \sigma_{4L} \}$. 
We divide the Hamiltonian $\hat{\mathcal{H}}$ into four blocks 
\begin{equation}
\hat{\mathcal{H}}^{(n)} = \sum_{\{ \sigma \} \{ \sigma' \}}\Tr
\Big[ \underbrace{ O^{(n)}_{\sigma_{\rm a} \sigma'_{\rm a}} O^{(n)}_{\sigma_{\rm b} \sigma'_{\rm b}} }_{\rm System} 
        \underbrace{ O^{(n)}_{\sigma_{\rm c} \sigma'_{\rm c}} O^{(n)}_{\sigma_{\rm d} \sigma'_{\rm d}} }_{\rm Environment} 
\Big] \ketbra{\sigma}{\sigma'}~,
\end{equation}
where $O^{(n)}_{\sigma_{\rm a} \sigma'_{\rm a}} \equiv {\cal W}_{\sigma_1\sigma'_1} \cdots{\cal  W}_{\sigma_L \sigma'_L}$ and so on. 
Here, we also introduce composite spin indexes $\sigma_{\rm a} = \{ \sigma_1, \cdots, \sigma_L \}$, $\sigma_{\rm b} = \{ \sigma_{L+1}, \cdots, \sigma_{2L} \}$, $\sigma_{\rm c} = \{ \sigma_{2L+1}, \cdots, \sigma_{3L} \}$, and $\sigma_{\rm d} = \{ \sigma_{3L+1}, \cdots, \sigma_{4L} \}$. 

Suppose that the groundstate eigenvector of this Hamiltonian, $\Psi^{(n)}_{\sigma_{\rm a}\sigma_{\rm b}\sigma_{\rm c}\sigma_{\rm d}}$, is calculated by the exact diagonalization like the Lanczos algorithm. 
Here, we remark that the relation between the Hamiltonian and the wavefunction $\Psi$ is quite reminiscent of Eq. (\ref{tm_eq}) for the 2D vertex model. 
Thus, $\Psi$ in the quantum systems approximately corresponds to $\rho$ in the HOTRG for 2D classical vertex model.

We divide the groundstate wave function of the total system into two blocks containing $2L$ sites.
Assuming the parity symmetry, we then perform SVD as $\Psi^{(n)}_{\sigma_{\rm a}\sigma_{\rm b}\sigma_{\rm c}\sigma_{\rm d}} = \sum_{\mu} \mathcal{U}^{(n)}_{\sigma_{\rm a} \sigma_{\rm b}, \mu} \Gamma_{\mu} \mathcal{U}^{(n)\dagger}_{\mu, \sigma_{\rm c} \sigma_{\rm d}}$, where $\mathcal{U}^{(n)}$ is an unitary matrix containing singular vectors and $\Gamma$ is a diagonal matrix containing nonnegative singular values.
We can use $\mathcal{U}^{(n)}$ as the RG transformation in the spatial direction for the Hamiltonian, namely
\begin{equation}
O^{(n+1)}_{\nu_{\rm a} \nu'_{\rm a}} = \sum_{\sigma_{\rm a}\sigma_{\rm b}\sigma'_{\rm a}\sigma'_{\rm b}} 
\mathcal{U}^{(n)\dagger}_{\nu_{\rm a}, \sigma_{\rm a}\sigma_{\rm b}}
O^{(n)}_{\sigma_{\rm a} \sigma'_{\rm a}} O^{(n)}_{\sigma_{\rm b} \sigma'_{\rm b}}
\mathcal{U}^{(n)}_{\sigma'_{\rm a}\sigma'_{\rm b},\nu'_{\rm a}}~.
\end{equation}
Thus, we can formulate a recursive numerical RG algorithm for the 1D quantum system similar to HOTRG.
Repeating  iterations,  we obtain the singular values at the fixed point, which we refer to as $\Gamma^{*}_\mu$. 

Particularly for the integrable model, the eigenvector of the Hamiltonian and the corresponding transfer matrix are exactly equivalent.
Indeed, we have performed a numerical computation of the above tensor RG for the 1D transverse-field Ising model in the off-critical region and actually confirmed the relation $\Gamma^{*}_{~} = \Omega^{*}$ under an appropriate normalization. 
At the critical point where the intrinsic correlation length is infinite, however, we should note that difference of geometries of $\Psi^{(n)}$ for the 1D quantum system and $\rho^{(n)}$ for the 2D classical model may affect the entanglement  structures of the effective fixed points, as in Sec. VI.



\begin{thebibliography}{99}
\bibitem{Kadanoff_1965_RG} L.~P.~Kadanoff, Physics {\bf 2}, 263 (1965). 
\bibitem{Wilson_1974_RG} E. Efrati, Z. Wang, A. Kolan, and L.P. Kadanoff, arXiv:1301.6323.
\bibitem{book:Burkhardt_1982_RSRG} T.W.~Burkhardt and J.M.J.~van~Leeuwen, {\it Real-Space Renormalization}, Topics in Current Physics Vol. 30 (Springer, Berlin, 1982).
\bibitem{White_1992_DMRG_1993_DMRG} S.R.~White, Phys. Rev. Lett. {\bf 69}, 2863 (1992); Phys. Rev. B {\bf 48},  10345 (1993).
\bibitem{book:Peschel_1999_DMRG} I.~Peschel, X.~Wang, M.~Kaulke, and K.~Hallberg, {\it Density Matrix Renormalization, A New Numerical Method in Physics}, (Springer, Berlin, 1999). 
\bibitem{Nishino_1999_DMRG} T.~Nishino, T.~Hikihara, K.~Okunishi, and Y.~Hieida, Int. J. Mod. Phys. B {\bf 13}, 1 (1999). 
\bibitem{Schollwock_2005_DMRG_Rev} U.~Schollw\"{o}ck, Rev. Mod. Phys. {\bf 77}, 259 (2005). 
\bibitem{Schollwock_2011_MPS_Rev} U.~Schollw\"{o}ck, Ann. Phys. (NY) {\bf 326}, 96 (2011). 
\bibitem{Xiang_2001_2D_DMRG} T.~Xiang, J.~Lou, and Z.~Su, Phys. Rev. B {\bf 64}, 104414 (2001).
\bibitem{Legeza_2003_k_space_DMRG} \"O.~Legeza and J.~S\'olyom, Phys. Rev. B {\bf 68}, 195116 (2003).
\bibitem{Legeza_2004_QI_and_DMRG} \"O.~Legeza and J.~S\'olyom, Phys. Rev. B {\bf 70}, 205118 (2004).
\bibitem{Vidal_2003_TEBD} G.~Vidal, Phys. Rev. Lett. {\bf 91}, 147902 (2003).
\bibitem{book:Baxter_1982_CTM} R.J.~Baxter, {\it Exactly Solved Models in Statistical Mechanics} (Academic Press, London, 1982). 
\bibitem{Peschel_1999_DM_spectra} I.~Peschel, M.~Kaulke, and \"O.~Legeza, Ann. Phys. {\bf 8}, 153 (1999). 
\bibitem{Okunishi_1999_DM_spectra} K.~Okunishi, Y.~Hieida, and Y.~Akutsu, Phys. Rev. E {\bf 59}, R6227 (1999). 
\bibitem{Nishino_1998_CTTRG} T.~Nishino and K.~Okunishi, J. Phys. Soc. Jpn. {\bf 67}, 3066 (1998).
\bibitem{Orus_2012} R.~Orus, Phys. Rev. B {\bf 85}, 205117 (2012).
\bibitem{Nishino_2000_TPVA} T.~Nishino, K.~Okunishi, Y.~Hieida, N.~Maeshima, and Y.~Akutsu, Nucl. Phys. B {\bf 575}, 504 (2000).
\bibitem{Nishino_2001_TPVA} T.~Nishino, Y.~Hieida, K.~Okunishi, N.~Maeshima and Y.~Akutsu, Prog. Theor. Phys. {\bf 105}, 409 (2001).
\bibitem{Verstraete_2006_PEPS} F.~Verstraete, M.M.~Wolf, D.~Perez-Garcia, and J.I.~Cirac, Phys. Rev. Lett. {\bf 96}, 220601 (2006).
\bibitem{Jordan_2009_MPS_iPEPS} J.~Jordan, R.~Or\'us, G.~Vidal, F.~Verstraete, and J.~I.~Cirac, Phys.
Rev. Lett. {\bf 101}, 250602 (2008).
\bibitem{Orus_2009_CTM_iPEPS} R.~Or\'us and G.~Vidal, Phys. Rev. B {\bf 80}, 094403 (2009).
\bibitem{Corboz_2010_CTM_ifPEPS} P.~Corboz, R.~Or\'us, B.~Bauer, and G.~Vidal, Phys. Rev. B {\bf 81}, 165104 (2010).
\bibitem{Orus_2012_review} R.~Or\'us , Phys. Rev. B {\bf 85}, 205117 (2012). 
\bibitem{Levin_2007_TRG} M.~Levin and C.P.~Nave, Phys.~Rev~Lett. {\bf 99}, 120601 (2007). 
\bibitem{Lathauwer_2000_HOSVD} L.~De~Lathauwer, B.~De~Moor, and J.~Vandewalle, SIAM J. Matrix Anal. Appl. {\bf 21}, 1253 (2000). 
\bibitem{Xie_2012_HOTRG} Z.Y.~Xie, J.~Chen, M.P.~Qin, J.W.~Zhu, L.P.~Yang, and T.~Xiang, Phys. Rev. B {\bf 86}, 045139 (2012). 
\bibitem{Xie_2009_SRG} Z.Y.~Xie, H.C.~Jiang, Q.N.~Chen, Z.Y.~Weng, and T.~Xiang, Phys. Rev. Lett. {\bf 103}, 160601 (2009).
\bibitem{Zhao_2010_SRG} H.H.~Zhao, Z.Y.~Xie, Q.N.~Chen, Z.C.~Wei, J.W.~Cai, and T.~Xiang, Phys. Rev. B {\bf 81}, 174411 (2010).
\bibitem{Suzuki_1976_trotter} M.~Suzuki, Prog. Theor. Phys. {\bf 56}, 1454 (1976).
\bibitem{Gu_2009_TEFR} Z.C.~Gu and X.G.~Wen, Phys. Rev. B {\bf 80}, 155131 (2009). 
\bibitem{ent_spec} This terminology of the entanglement spectrum is different from that for quantum systems, although the implication of them is essentially the same. The relation between $\Omega_\mu$ at the RG fixed point and the entanglement spectrum for 1D quantum systems $\epsilon_\mu$ is given by $\epsilon_\mu = -\log\Omega^2_\mu$. 
\bibitem{Nishino_2013_boundary_mps} T. Nishino, in preparation. 
\bibitem{Tsang_1979_2DIsing} S.~K.~Tsang, J. Stat. Phys. {\bf 20}, 95 (1979). 
%
\bibitem{duality} The explicit form of the sequence $c_n$ has two possibilities, depending on the construction of CTM. If the corner of CTM is located at the center of the plaquette, $c_n$ is given by Eq.~(\ref{sequence}). While the corner is located at the spin site, $c_n$ is given by the dual of Eq.~(\ref{sequence}) with the dual temperature.
%
\bibitem{Onsager_1944_2DIsing} L.~Onsager, Phys. Rev. {\bf 65}, 117 (1944). 
%
\bibitem{symmetric} For $D = 49$, the renormalized vertex is decomposed to a $7 \times 7$ CTMs. 
When $\sqrt{D}$ is not an integer, the dimension of the CTM becomes  $[ \sqrt{D} ]$,  where $[\sqrt{D} ]$ is the greatest integer not greater than $\sqrt{D}$.
We also note that, sometimes, a vertex may be decomposed into CTMs of different dimensions, e.g.  for $D = 48$, we have CTMs of $6 \times 6$ and $8 \times 8$.
\bibitem{Verstraete_2004_MPDO} F.~Verstraete, J.~J.~Garcia-Ripoll, and J.~I.~Cirac,  Phys. Rev. Lett. {\bf 93}, 207204 (2004).
\bibitem{McCulloch_2007_MPO} I.~P.~McCulloch, J. Stat. Mech.: Theor. Exp. P10014 (2007).
\bibitem{Tagliacozzo_2008_EE_MPS} L.~Tagliacozzo, T.~R.~de~Oliveira, S.~Iblisdir, and J.I.~Latorre, Phys. Rev. B {\bf 78}, 024410 (2008). 
\bibitem{Pollmann_2009_FES} F.~Pollmann, S.~Mukerjee, A.~M.~Turner, and J.~E.~Moore, Phys. Rev. Lett. {\bf 102}, 255701 (2009). 
\bibitem{Pirvu_2012_FSS-FES} B.~ Pirvu, G.~Vidal, F.~Verstraete, and L.~Tagliacozzo, Phys. Rev. B {\bf 86}, 075117 (2012).  
\bibitem{Holzhey_1994_EE_CFT} C.~Holzhey, F.~Larsen, and F.~Wilczek, Nucl. Phys. {\bf B424}, 443 (1994).
\bibitem{Osterloh_2002_EE} A.~Osterloh, L.~Amico, G.~Falci, and R.~Fazio, Nature (London) {\bf 416}, 608 (2002).
\bibitem{Vidal_2003_EE} G.~Vidal, J.~I.~Latorre, E.~Rico, and A.~Kitaev, Phys. Rev. Lett. {\bf 90}, 227902 (2003).
\bibitem{Calabrese_2004_EE_CFT} P. Calabrese and J. Cardy, J. Stat. Mech. P06002, (2004).
\bibitem{Refael_2004_EE_CFT} G. Refael and J.~E.~Moore, Phys. Rev. Lett. 93, 260602 (2004).
\bibitem{Amico_2008_EE_refiew} L.~Amico, R.~Fazio, A.~Osterloh, and V.~Vedral, Rev. Mod. Phys. {\bf 80}, 517 (2008).

\bibitem{Nishino_1996_CTMRG} T.~Nishino and K.~Okunishi, J. Phys. Soc. Jpn. {\bf 65}, 891 (1996).
\bibitem{Nishino_1996_CTMRG_scaling} T.~Nishino, K.~Okunishi and M.~Kikuchi, 
Phys. Lett. A {\bf 213}, 69 (1996). 

\end{thebibliography}
\end{document}